\documentclass[conference,a4paper]{IEEEtran}
\IEEEoverridecommandlockouts
\usepackage{cite}
\usepackage{amsmath,amssymb,amsfonts,amsbsy}
\usepackage{algorithm}
\usepackage[noend]{algpseudocode}
\usepackage{graphicx}
\usepackage{textcomp}
\usepackage[table]{xcolor}    
\usepackage{cases}
\usepackage{stfloats}
\usepackage[letterspace=-125]{microtype}

\usepackage[LGR,T1]{fontenc}

\DeclareMathAlphabet{\mathgtt}{LGR}{cmtt}{m}{n}

\usepackage{booktabs}
\usepackage[export]{adjustbox}
\usepackage{multirow}
\usepackage{enumitem}
\usepackage{titlesec}

\usepackage{diagbox}

\usepackage{stackengine}

\renewcommand{\baselinestretch}{0.975}
\usepackage[font=footnotesize,skip=10pt]{caption}

\titlespacing{\subsection}{1em}{\parskip}{-\parskip}
\titlespacing{\subsubsection}{1em}{\parskip}{-\parskip}

\newcommand{\mtt}[1]{$\mathtt{#1}$}

\newcommand{\ckmtt}[1]{\textls[-50]{\scalebox{0.95}[1.0]{$\mathtt{#1}$}}}

\newcommand{\mC}{\mathcal{C}}

\newcommand\mydots{\ifmmode\ldots\else\makebox[1em][c]{.\hfil.\hfil.}\thinspace\fi}

\def\BibTeX{{\rm B\kern-.05em{\sc i\kern-.025em b}\kern-.08em
    T\kern-.1667em\lower.7ex\hbox{E}\kern-.125emX}}
\begin{document}

\title{
\adjustbox{width=\textwidth}{Unleashing the Power of T1-cells in SFQ Arithmetic Circuits}
}

\author{
\IEEEauthorblockN{Rassul Bairamkulov, Mingfei Yu and Giovanni De Micheli}
\IEEEauthorblockA{Integrated Systems Laboratory, EPFL} 
\IEEEauthorblockA{Lausanne, Switzerland}
\IEEEauthorblockA{\{rassul.bairamkulov,mingfei.yu,giovanni.demicheli\}@epfl.ch}
\vspace{-0.8cm}
}

\maketitle

\begin{abstract}

Rapid single-flux quantum (RSFQ), a leading cryogenic superconductive electronics (SCE) technology, offers extremely low power dissipation and high speed.
However, implementing RSFQ systems at VLSI complexity faces challenges, such as substantial area overhead from gate-level pipelining and path balancing, exacerbated by RSFQ's limited layout density.

T1 flip-flop (T1-FF) is an RSFQ logic cell operating as a pulse counter.
Using T1-FF the full adder function can be realized with only 40\% of the area required by the conventional realization.
This cell however imposes complex constraints on input signal timing, complicating its use.
Multiphase clocking has been recently proposed to alleviate gate-level pipelining overhead. 
The fanin signals can be efficiently controlled using multiphase clocking.
We present the novel two-stage SFQ technology mapping methodology supporting the T1-FF. 
Compatible parts of the SFQ network are first replaced by the efficient T1-FFs.
Multiphase retiming is next applied to assign clock phases to each logic gate and insert DFFs to satisfy the input timing.
Using our flow, the area of the SFQ networks is reduced, on average, by 6\% with up to 25\% reduction in optimizing the 128-bit adder.

\end{abstract}





\begin{figure*}[b!]
\vspace{-1em}
\hrule
\includegraphics[trim={5mm 0 0 4mm},clip,scale=0.725,valign=b]{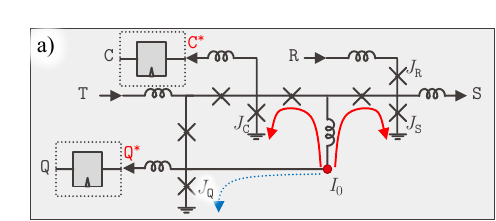} 
\includegraphics[trim={5mm 0 0 4mm},clip,scale=0.775,valign=b]{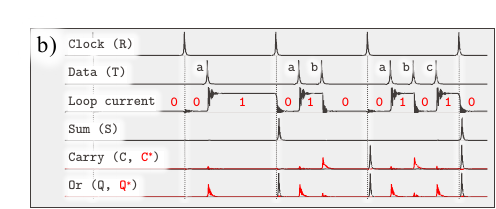}
\begin{minipage}[t][0pt][b]{18em}
\includegraphics[trim={5mm 0 0 4mm},clip,scale=0.77,valign=b]{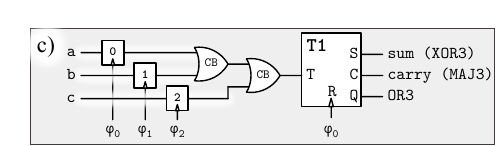}
\caption{\scalebox{0.9}{T1-FF. a) Circuit, b) simulation, c) full adder.}}\label{fig:t1_cell}
\end{minipage}
\end{figure*}

\section{Introduction}

Rapid Single-Flux Quantum (RSFQ)\cite{likharev1985} is a prominent superconductive digital technology.
RSFQ circuits operate at tens to hundreds of gigahertz and dissipate two to three orders of magnitude less power as compared to CMOS, even accounting for refrigeration \cite{krylov2022book}.
These advantages position RSFQ as a compelling candidate for large-scale stationary computing~\cite{holmes2013energy}, space electronics~\cite{krylov2022book} and interface circuitry for quantum computing systems~\cite{jokar2022}.
These applications require efficient arithmetic circuits.


RSFQ systems consist of Josephson Junctions (JJ) and superconductive storage loops communicating using single flux quantum (SFQ) pulses.
\footnote{For background in RSFQ technology, an interested reader is referred to \cite{krylov2022book}}
Due to pulse-based operation, most SFQ gates, such as \mtt{NOT} and \mtt{XOR}, require a clock signal, necessitating gate-level pipelining.
Fanins to each clocked SFQ gate need to have equal logic depth to operate correctly.
To satisfy this requirement, dummy DFFs are inserted into the network,  occupying a significant portion of the layout.

\subsection{T1-flip-flop}

T1-flip-flop~\cite{polonsky_bff} is an SFQ device with two inputs, \mtt{T} (\textit{toggle}) and  \mtt{R} (\textit{reset}); and three outputs, \mtt{S} (\textit{sum}), \mtt{C} (\textit{carry}) and \mtt{Q}, as illustrated in Fig. \ref{fig:t1_cell}a.
Initially, the bias current $I_0$ flows along the blue dotted line, corresponding to the internal state \mtt{0}.
A pulse arriving at input \mtt{T} switches $J_{\mathtt{Q}}$, producing the pulse at output $\mathtt{Q}^*$ (see Fig.~\ref{fig:t1_cell}b).
The bias current is redirected along the red solid arrows, corresponding to the storage of logical \mtt{1}.
A second pulse at input \mtt{T} switches $J_{\mathtt{C}}$, producing the pulse at output $\mathtt{C}^*$, and resetting the bias current towards $J_{\mathtt{Q}}$, i.e., logical \mtt{0}.
If the loop state is \mtt{1}, a pulse at the input \mtt{R} switches $J_{\mathtt{S}}$, producing the pulse at output $\mathtt{S}$, while resetting the loop state to \mtt{0}.
If the loop state is \mtt{0}, a pulse at the input \mtt{R} is simply rejected by $J_{\mathtt{R}}$.

The T1-FF can realize a full adder with only 29 JJs, 60\% fewer than a regular implementation \cite{yorozu2002} (see Fig.~\ref{fig:t1_cell}c).
Outputs $\mathtt{R}$, $\mathtt{C}$, and $\mathtt{Q}$ execute, respectively, \mtt{XOR3}, majority-3 (\mtt{MAJ3}), and \mtt{OR3} functions.
In addition, outputs $\mathtt{C}^*$ and $\mathtt{Q}^*$ can be connected to inverters to produce inverted \mtt{MAJ3} and \mtt{OR3}.
Therefore, the extended T1-FF can efficiently produce up to five synchronous outputs. 
The main challenge of using T1-FFs is temporal separation of input pulses. 
Two overlapping input pulses may be treated as a single pulse, producing a data hazard. 
We propose using multiphase clocking to mitigate this issue.

\subsection{Multiphase clocking} \label{sec:multiphase}

An $n$-phase system utilizes $n$ periodic signals $\{t_0,\cdots,t_{n-1}\}$ operating at the same frequency \cite{li_beerel}. 
Each clocked element $g$ within the network is synchronized by only one clock signal at phase $\varphi(g)$. 
The epoch $S(g)$ of a gate $g$ is defined as the number of clock cycles separating the gate $g$ from the primary inputs.
The clock signals are ordered by phase $\varphi\in\{0,\cdots,n-1\}$, i.e., during any epoch, the clock signal $t_i$ arrives before clock signal $t_j$ if $i<j$.
For convenience, we define a \emph{stage} $\sigma(g)$ of a gate $g$ as 
\begin{equation}
    \sigma(g)=nS(g)+\varphi(g).
\end{equation}

We observe that multiphase clocking enables precise control of the input arrival time, as illustrated by phases $\mathgtt{f_0}$, $\mathgtt{f_1}$, and $\mathgtt{f_2}$ in Fig.~\ref{fig:t1_cell}c.
The input $a$ is released to the T1-FF at $\mathgtt{f_0}$, next $b$ is released at $\mathgtt{f_1}$, and, finally, $c$ is released at $\mathgtt{f_2}$;
i.e., assigning three different phases to the inputs of a T1-FF is sufficient to ensure no temporal overlap.

\section{T1-aware technology mapping}\label{sec:fa_detection}

We present next the three-stage T1-aware technology mapping flow.
First, compatible parts of the logic network are replaced by the T1-FF.
Then, we formulate an integer linear programming problem to assign a phase to each gate, while minimizing the number of DFFs.
Finally, the DFFs are inserted to satisfy the timing requirements of each gate, including T1-FFs.


\subsection{T1-FF detection}


Our T1-FF detection is based on cut enumeration \cite{cong1999cut} followed by Boolean matching \cite{DeMicheli1994book}.
If a set of cuts $\mC=\{C(u_1),\dots,C(u_n)\}$, $2\leq n\leq 5$ sharing the same leaves $\{a,b,c\}$ executes the functions implementable with the T1-FF, the cuts $\{C(u_1),\dots,C(u_n)\}$ are considered for being replaced by a T1-FF. 
To ensure the substitution is beneficial, the area reduction $\Delta A$ due to replacement is calculated as 
\begin{equation}\label{eq:area_gain}
\Delta A = \sum\limits_{i=1}^n A(\ckmtt{MFFC}(u_i)) - A_{\ckmtt{T1}}(\mC),
\end{equation}
where $A(\ckmtt{MFFC}(u_i))$ is the total area of the nodes within the maximum fanout free cone (MFFC) of node $u_i$, and $A_{\ckmtt{T1}}(\mC)$ is the area of the T1-FF implementing the functions realized by cuts in $\mC$ considering possible input and output negations. 
If $\Delta A > 0$, the MFFCs of nodes $u_1,\dots,u_n$ are replaced by the T1-FF-based circuit.

\subsection{Phase assignment}\label{sec:phase_assignment}

The phase assignment closely follows the integer linear programming (ILP) procedure described in \cite{bairamkulov_aspdac24}.
The T1-FF however requires the constraint and objective functions to be modified.
The clock stage $\sigma_{\ckmtt{T1}}$ of T1-FF is constrained as
\begin{equation} \label{eq:phase_assignment_4}
    \sigma \left( j \right) \geq \max \left( 
    \sigma \left( i_1 \right) + 3,
    \sigma \left( i_2 \right) + 2,
    \sigma \left( i_3 \right) + 1 \right),
\end{equation}
where $i_1, i_2, i_3$ are the fanins of T1-FF and $\sigma(i_1)\leq\sigma(i_2)\leq\sigma(i_3)$.
Condition (\ref{eq:phase_assignment_4}) is incorporated into the set of ILP constraints, ensuring that the necessary DFFs can be inserted between the inputs and the T1-FF. 
The number of DFFs required by a T1-FF is determined as 
\begin{equation} \label{eq:cost}
    \begin{split}   
    c_{\ckmtt{T1}} = & \left( \phi(i_1)=\phi(i_2) \right) \wedge \left( \sigma_{\ckmtt{T1}} - \sigma(i_1) \leq n \right) + \\ 
    + & \left( \phi(i_2)=\phi(i_3) \right) \wedge \left( \sigma_{\ckmtt{T1}} - \sigma(i_2) \leq n \right).
    \end{split} 
\end{equation}
$c_{\ckmtt{T1}}$ is added to the ILP objective function. 

\subsection{DFF insertion}\label{sec:dff_insertion}

After assigning the stage to each gate, the DFFs can be inserted to each datapath.
We extend the DFF insertion methodology based on CP-SAT, described in \cite{bairamkulov_aspdac24}, to support the T1-FF. 
The inputs to the T1-FF should arrive at different stages. 
To satisfy this condition, the DFFs  $d_1$, $d_2$, $d_3$ preceding the T1-FF at stage $\sigma_{\ckmtt{T1}}$ should be placed at a different stage,
\begin{equation}\label{eq:condition_5}
\sigma(a) \neq \sigma(b) \text{\hspace{0.3em}} \forall \text{\hspace{0.3em}} a\neq b \text{\hspace{1em}} a,b \in \{d_1, d_2, d_3\}
\end{equation}

\begin{table*}[t]
\centering
\caption{Multiphase clocking with T1 cells applied to a subset of EPFL and ISCAS benchmark circuits }
\label{tab:results}
\begin{tabular}{l||rr||rrr|rr||rrr|rr||rrr|rr}
\toprule
 & \multicolumn{2}{c||}{T1 cells} & \multicolumn{3}{c|}{\#DFF} & \multicolumn{2}{c||}{Ratio vs.} & \multicolumn{3}{c|}{Area} & \multicolumn{2}{c||}{Ratio vs.} & \multicolumn{3}{c|}{Depth} & \multicolumn{2}{c}{Ratio vs.}\\
benchmark & found & used & 1$\mathgtt{f}$ & 4$\mathgtt{f}$ & T1 & 1$\mathgtt{f}$ & 4$\mathgtt{f}$ & 1$\mathgtt{f}$ & 4$\mathgtt{f}$ & T1 & 1$\mathgtt{f}$ & 4$\mathgtt{f}$ & 1$\mathgtt{f}$ & 4$\mathgtt{f}$ & T1 & 1$\mathgtt{f}$ & 4$\mathgtt{f}$\\
\cmidrule(rl){1-1} \cmidrule(rl){2-3}  \cmidrule(rl){4-6}  \cmidrule(rl){7-8} \cmidrule(rl){9-11}  \cmidrule(rl){12-13} \cmidrule(rl){14-16}  \cmidrule(rl){17-18} 
\mtt{adder} & 127	&	127 & 32'768 & 7'963 & 5'958 & 0.18 & 0.75 & 238'419 & 64'784 & 48'844 & 0.20 & 0.75 & 128 & 32 & 33 & 0.26 & 1.03 \\
\rowcolor{gray!15}
\mtt{c7552} & 17	&	9 & 2'489 & 713 & 765 & 0.31 & 1.07 & 32'038 & 19'606 & 19'907 & 0.62 & 1.02 & 16 & 4 & 5 & 0.31 & 1.25 \\
\mtt{c6288} & 142	&	142 & 2'625 & 1'431 & 1'349 & 0.51 & 0.94 & 47'198 & 38'840 & 35'386 & 0.75 & 0.91 & 29 & 8 & 10 & 0.34 & 1.25 \\
\rowcolor{gray!15}
\mtt{sin} & 81	&	77 & 13'416 & 4'631 & 4'714 & 0.35 & 1.02 & 164'938 & 103'443 & 102'806 & 0.62 & 0.99 & 88 & 22 & 25 & 0.28 & 1.14 \\
\mtt{voter} & 252	&	252 & 10'651 & 5'779 & 5'584 & 0.52 & 0.97 & 222'101 & 187'997 & 182'972 & 0.82 & 0.97 & 38 & 10 & 11 & 0.29 & 1.10 \\
\rowcolor{gray!15}
\mtt{square} & 861	&	806 & 44'675 & 16'645 & 14'304 & 0.32 & 0.86 & 525'311 & 329'101 & 301'287 & 0.57 & 0.92 & 126 & 32 & 32 & 0.25 & 1.00 \\
\mtt{multiplier} & 824	&	769 & 58'717 & 14'641 & 13'745 & 0.23 & 0.94 & 682'792 & 374'260 & 356'984 & 0.52 & 0.95 & 136 & 33 & 36 & 0.26 & 1.09 \\
\rowcolor{gray!15}
\mtt{log2} & 644	&	593 & 86'985 & 33'790 & 33'946 & 0.39 & 1.00 & 978'178 & 605'813 & 598'292 & 0.61 & 0.99 & 160 & 40 & 47 & 0.29 & 1.18 \\
\cmidrule(rl){1-1} \cmidrule(rl){2-3}  \cmidrule(rl){4-6}  \cmidrule(rl){7-8} \cmidrule(rl){9-11}  \cmidrule(rl){12-13} \cmidrule(rl){14-16}  \cmidrule(rl){17-18} 
\rowcolor{gray!30}
\multicolumn{6}{l}{Average}& 0.35 & \multicolumn{1}{l}{0.94} & \multicolumn{3}{l}{} & 0.59 & \multicolumn{1}{l}{0.94} & \multicolumn{3}{l}{} & 0.29 & 1.13\\
\bottomrule
\end{tabular}
\end{table*}

\vspace{-0.3em}
\section{Experimental results}\label{sec:results}
\vspace{-0.3em}

We integrate the proposed methodology into the technology mapping flow implemented in \mtt{mockturtle} logic synthesis library \cite{soeken2018epfl}.
Phase assignment and DFF insertion procedures are implemented using Google OR-Tools~\cite{ortools}.
We apply our flow to synthesize a subset of EPFL~\cite{soeken2018epfl} and ISCAS~\cite{hansen1999unveiling} benchmark circuits implementing arithmetic functions. 
We ran our experiments on a laptop with an Apple M1 10-core CPU with 64 GB of RAM. 
The number of path-balancing DFFs, circuit area (expressed in the number of JJs), and logic depth (in cycles) are shown in Table~\ref{tab:results}.
We compare our synthesis results (column T1) with the single-phase (1$\mathgtt{f}$) and four-phase (4$\mathgtt{f}$) clocking without T1-FF.
Compared to 4$\mathgtt{f}$, our methodology achieves, on average, a 6\% improvement in both area and the number of path-balancing DFFs at the cost of a 13\% increase in the logic depth. 
The increase in depth can be attributed to the additional stages required by the T1-FFs.
The largest reduction is observed in the \mtt{adder} circuit where almost the entire circuit is replaced with the T1-FFs, yielding a 25\% improvement in area. 
Significant reduction is also observed in \mtt{voter}, \mtt{square} and \mtt{multiplier}, while \mtt{c7552} and \mtt{sin} yielded inferior area, likely due to the increase in the circuit depth, requiring additional path balancing DFFs.

\bibliographystyle{sty/myIEEE.bst}
\vspace{-0.3em}

\vspace{-0.3em}

\end{document}